\documentstyle[preprint,aps,epsf]{revtex}
\tighten
\begin{document}
\draft
\preprint{}
\title{Monopole Black Hole Skyrmions}
\author{I.~G.~Moss$^{1}$\thanks{email: {\tt ian.moss@ncl.ac.uk}}
            ,N.~Shiiki$^{1}$
             \thanks{email: 
            {\tt noriko.shiiki@ncl.ac.uk}} and E.~Winstanley$^{1,2} 
            $\thanks{email: 
            {\tt elizabeth.winstanley@oriel.ox.ac.uk }}}
\address{$1$.
Department of Physics, University of Newcastle Upon Tyne, NE1 7RU U.K.}
\address{$2$.Department of Physics (Theoretical Physics), 1 Keble Road, 
Oxford, OX1 3NP U.K.}
\date{May 2000}
\maketitle
\begin{abstract}
Charged black hole solutions with pion hair are discussed. 
These can be used to study monopole black hole catalysis 
of proton decay. There also exist multi-black hole skyrmion 
solutions with BPS monopole behaviour.\\ 
\end{abstract}
\pacs{Pacs numbers: 0470D, 0420J, 1239D}
\narrowtext

\section{Introduction}

It is known that baryons can emerge as solitons in an effective meson field theory. In these models the baryon number is identified with a topological charge. Since this identification was originally made by T.H.R.Skyrme $40$ 
years ago \cite{skyrme}, the soliton solution is called the skyrmion.  

Skyrmion models provide a useful way of analysing the monopole catalysis of proton decay, as originally pointed out by Callan and Witten \cite{callan}. The decay process can be allowed because, in the presence of a monopole, the baryon number is no longer equal to the topological charge of the meson field. In fact, there exist non-topological solitons with non-zero baryon number which can decay without topological problems.

Skyrmions have also been used to examine the absorbtion of baryons by microscopic black holes \cite{moss,luckock,zurich}. Solutions representing skyrmions partially absorbed by Schwarzschild black holes provide a semiclassical framework to study the absorption rate of a proton by a black hole of comparable size. However, this process is rather insignificant because black holes of the size of a proton have large fluxes of Hawking radiation which swamp the proton decay.  

In this paper we analyse black hole solutions with skyrmion hair and magnetic charge, providing the semiclassical framework in which to study monopole black hole catalysis of proton decay. The flux of Hawking radiation from these holes is less than ordinary holes and even vanishes in extremal cases \cite{gibbons}. The black hole mass for abelian monopoles has the lower bound
\begin{equation}
	M=\frac{p}{\sqrt{G}}=pm_{{\rm pl}}\approx 2.54\times 10^{-7} Kg  \; ,
	{}
\end{equation}
where ${\rm e}$ is a rationalised electric charge in gaussian units 
determined from the Dirac quantisation condition $({\rm e}p=1)$ 
as $p\approx 11.7$ and $m_{{\rm pl}}$ is 
the planck mass which is approximately $ 2.1768 \times 10^{-8} Kg$.  
This kind of small black hole may be considered to have been created 
in the early universe and remain as a stable relic today. More complicated monopole solutions are also possible in Grand Unified Theories, depending on the details of the Higgs sector \cite{higgs}, but we will restrict attention to the simplest case.

In the extreme case, $M=pm_{\rm pl}$, we find the remarkable situation that multiple black hole solutions are possible in which the gravitational, electromagnetic, and strong forces between the monopoles are all in balance. In this respect the solitons behave in an analagous way to BPS monopole solutions in the Yang-Mills-Higgs system \cite{bps}.

We find that the non-topological skyrmion solutions are stable within the confines of our model. This is, in part, due to the fact that we have not included the electron. The black hole cannot swallow the proton whole because this would tip it over the extremal limit. Light, charged particles are therefore needed to carry away the proton's charge when it decays. We discuss a way of including this effect in the conclusion.

\section{The lagrangian}
 
We shall consider models with a charged $SU(2)$ meson field $U$. The Lagrangian is based on a gauged version of the original skyrmion Lagrangian \cite{skyrme}. It should be regarded as a minimal form of the meson effective action, since extra terms could also be included. However, some of the terms in the Lagrangian which couple the chiral field to electromagnetism can be deduced from current algebra techniques and were constructed by Callan and Witten \cite{callan}.

The Lagrangian can be divided into into four parts,
\begin{equation}
	{\cal L} = {\cal L}_{1} + {\cal L}_{2} + {\cal L}_{3} + {\cal L}_{4}
	\label{lag}
\end{equation}
where the meson parts are  
 \begin{eqnarray*}
	{\cal L}_{1} &= &\frac{F_{\pi}^{2}}{16}{\rm 
	tr}\left(U^{-1}D_{\mu}UU^{-1}D_{\mu}U\right)+\frac{1}
        {32a^{2}}{\rm tr}\left([U^{-1}D_{\mu}U,U^{-1}D_{\mu}U]\right)\\ 
	{\cal L}_{2} & = & 
	 \frac{{\rm e}}{16\pi^{2}}\epsilon^{\mu\nu\rho\sigma}A_{\mu}{\rm 
	 	 tr}\left\{Q\left(\partial_{\nu}UU^{-1}\partial_{\rho}UU^{-1}
	 	 \partial_{\sigma}UU^{-1}+U^{-1}\partial_{\nu}UU^{-1}
	 	 \partial_{\rho}UU^{-1}\partial_{\sigma}U\right)\right\}  \\
	 &+ &\frac{i{\rm e}^{2}}{8\pi^{2}}\epsilon^{\mu\nu\rho\sigma}
	 (\partial_{\mu}A_{\nu})A_{\rho}{\rm 
	 tr}\left(Q^{2}\partial_{\sigma}UU^{-1}+Q^{2}U^{-1}
	 \partial_{\sigma}U+\frac{1}{2}Q\partial_{\sigma}UQU^{-1}-
	 \frac{1}{2}QUQ\partial_{\sigma}U^{-1}\right) 
  \end{eqnarray*}
and the free actions are
\begin{equation}
	{\cal L}_{3}  =  -\frac{1}{16\pi}F_{\mu\nu}F^{\mu\nu},
\quad   {\cal L}_{4}  =  \frac{1}{16\pi G}R.
\end{equation}
The quantity $Q$ is the charge matrix of quarks and $D$ is a covariant derivative defined by
\begin{equation}
	Q=\left(
	\begin{array}{cc}
		\frac{2}{3} & 0  \\
		0 & -\frac{1}{3}
	\end{array}
	{}\right)
\end{equation}
\begin{equation}
	D_{\mu}U = \partial_{\mu}U - i{\rm e}A_{\mu}[Q,U] .
	{}
\end{equation}
The abelian gauge field $A_{\mu}$ and electric charge ${\rm e}$ are in unrationalised units, and $\hbar = c = 1 $.

In the spherically symmetric case with a magnetic charge the gauge field has the form
\begin{equation}
	A=p(1-\cos \theta )d\phi + \Phi dt
	\label{dirac}
\end{equation}
where $p$ is a magnetic charge and the Dirac quantisation condition 
is $ p{\rm e} =1$. The usual Skyrmion has a magnetic moment which would interact with a magnetic monopole and break the spherical symmetry. We use instead a non-topological ansatz for the chiral field, 	
	\begin{equation}
		U=e^{if(r,t)\sigma_{3}}.
		\label{hedge}
	\end{equation}
Despite the fact that this field is made up of neutral pions and commutes with the charge matrix $Q$, it has a non-zero total electric and baryonic charge due to the effects of anomalies, as we shall see in the next section.

\section{Baryon Number and Electric Charge}

The gauge invariant baryon current was constructed by 
Callan and Witten
\begin{eqnarray}
	j_B^{\mu}  & = &  \frac{ \epsilon^{\mu \nu \alpha\beta }}
        {48\pi ^{2}}\left[{\rm 
     	tr}U^{-1}\partial_{\nu}UU^{-1}\partial_{\alpha}UU^{-1}
        \partial_{ \beta }U  \right.
     	\\
	 & &+  3i{\rm e}A_{\nu}{\rm tr}Q(U^{-1}
          \partial _{\alpha}UU^{-1}\partial 
     	_{\beta}U-\partial_{\alpha}UU^{-1}\partial_{\beta}UU^{-1})  \\
	 & &  \left. +  3i{\rm e}\partial_{\nu}A_{\alpha}{\rm 
      	tr}Q(U^{-1}\partial_{\beta}U+\partial_{\beta}UU^{-1})\right] 
          \; . \label{baryon}
\end{eqnarray}
Only the third term survives for the chiral field anzatz (\ref{hedge}), with a resulting baryon charge 
\begin{equation}
	n_B = {ep\over2\pi}\left[f(\infty)-f(0)\right] \; .
	{}
\end{equation}
The solution with the boundary conditions $f(0)=0$ 
and $f(\infty)=2\pi$ posesses unit baryon number and hence it can be 
interpreted as a baryon surrounding the monopole.

If the field anzatz (\ref{hedge}) is substituted into the meson and electromagnetic interaction terms in the Lagrangian, they become
\begin{eqnarray}
{\cal L}_1&=&-{F_\pi^2\over 8}(\partial f)^2,
\label{lag1}\\
{\cal L}_2&=&-{e^2\over 8\pi^2}E_i B^i\,f,\\
{\cal L}_3&=&{1\over 8\pi}\left(E^2-B^2\right),
\end{eqnarray}
where the index $i=1,2,3$ and the electromagnetic fields $E_i$ and $B_i$ are defined by $F_{0i}=(-g_{tt})^{1/2}E_i$ and $F_{ij}=\epsilon_{ijk}B^k$. When combined,
\begin{equation}
{\cal L}_2+{\cal L}_3={1\over 8\pi}\left(E-{e^2\over 2\pi}Bf\right)^2
-{1\over 8\pi}B^2\left(1-{e^4\over 4\pi^2}f^2\right)\label{lag2}
\end{equation}
The extrema of the action occur when the electric field is given by
\begin{equation}
E={e^2\over 2\pi}Bf
\end{equation}
This situation is reminisent of the factorisation of the Lagrangian that occurs for a BPS monopole \cite{bps}.

The electric field implies a total charge
\begin{equation}
q={e^2p\over 2\pi}f\label{charge}
\end{equation}
or asymptotically
\begin{equation}
q_\infty=n_Be
\end{equation}
and the $n_B=1$ solution can therefore be interpreted as a proton.

If a black hole appears in the background, the inner boundary 
condition for the field $f$ should be imposed not at the origin 
but at the event horizon $r=r_{+}$.
Thus the baryon number in the presence of an event horizon 
will be defined as 
\begin{equation}
	n_B = \frac{ep}{2\pi}\left[f(\infty)-f(r_{+})\right] \; .
	{}
\end{equation}
If $f(r_{+})=0$ the baryon number is still an integer and conserved. 
This configuration represents a proton tightly bound to the black hole. 
On the other hand if $f$ takes some positive value at the horizon 
the baryon number is not an integer and the skyrmion carries 
fractional baryon number and electric charge. 
This configuration will be interpreted as a proton 
partially swallowed by the black hole.
In particular, $f(r_{+})=2\pi$ means that the black hole 
has swallowed a whole proton leaving nothing outside 
the horizon. 

It is interesting to observe that,
while the baryon number disappears inside the horizon, 
the electric charge of the black hole can still be measured outside, 
turning the monopole black hole into a dyon black hole. 
Therefore, while the baryon number conservation is violated, charge 
conservation is not violated.

 \section{Extremal black hole solutions}

In the extremal case we can obtain a general solution based on the Papapetrou-Majumbar metrics \cite{maj,pap}. We begin with the background metric fixed and later generalise to solve the full Einstein equations with chiral matter. The Papapetrou Majumbar metrics have the form
\begin{equation}
ds^2=-U^{-2}dt^2+U^2(dx^2+dy^2+dz^2)\label{papmaj}
\end{equation}
where
\begin{equation}
U=1+\sum_{n=1}^{n_M}{GM_n\over R_n}
\end{equation}
and $R_n$ is the ordinary Euclidean distance from the point mass $M_n$
located in three dimensional space. We also associate these
point masses with magnetic charges $P_n=G^{1/2}M_n$, and the
magnetic field
\begin{equation}
B=G^{-1/2}U^{-1}\partial U.
\end{equation}
The matter lagrangian obtained earlier (\ref{lag1}) has the form
\begin{equation}
{\cal L}=-{1\over 8}F_\pi^2(\partial f)^2
-{1\over 8\pi}B^2\left(1+\alpha^2f^2\right)
\label{masslag}
\end{equation}
where we will set
\begin{equation}
\alpha={e^2\over 2\pi},\qquad\mu^2=\pi GF_\pi^2
\end{equation}
The skyrme field equation obtained from the Lagrangian on this background
becomes
\begin{equation}
-\mu^2\partial^2f+\alpha^2 U^{-2}(\partial U)^2f=0
\end{equation}
We loose no generality by taking equal charges $P_n=p$. The solution with baryon number $n_B=n_M$ is then
\begin{equation}
f=2\pi U^{-s}
\end{equation}
where
\begin{equation}
s=-{1\over 2}+\sqrt{{1\over 4}+{\alpha^2\over \mu^2}}
\end{equation}
Since $F_\pi\ll m_{pl}$, we can use $s\approx\alpha/\mu$ for the pion model.

For a single black hole, the Reissner-Nordstrom coordinate $r$ is related to $R$ by $R=r-r_+$ and we obtain
\begin{equation}
f=2\pi\left(1-{r_+\over r}\right)^s.\label{oneex}
\end{equation}
The field is effectively expelled by the black hole and vanishes on the horizon $r=r_+$.

The mass of the chiral field configuration can be obtained by integrating the Lagrangian (\ref{masslag}),
\begin{equation}
m_f={1\over 8}F_\pi^2\int_\Sigma f\partial_if dS^i
\end{equation}
where $\Sigma$ is a large surface containing all of the masses. This gives
\begin{equation}
m_f=2\pi^3sF_\pi^2G\sum_n M_n\approx
\pi^{3/2}n_BeF_\pi
\end{equation}
The total mass in the chiral field is much less than one baryon mass per mass point. We can see how it is energetically favourable for a free skyrmion to change its internal configuration from the original Skyrme form to the simpler form used here when it comes into contact with a black hole monopole. The topological description of this transformation for a single monopole is exactly as described in reference \cite{callan}.

It is interesting to see that the electrostatic energy cancells due to the factorisation occuring in the lagrangian (\ref{lag2}). The chiral field mass is independent of the separation of the holes and therefore there are no forces between then. This is similar to situation for BPS monopole solutions \cite{bps}, and suggests that there is a solution of the full Einstein-matter system. This existence of this solution will now be demonstrated.

The spatial part of the Einstein tensor for the matric (\ref{papmaj}) is
\begin{equation}
G_{ij}=-2U^{-2}(\partial_i U)(\partial_j U)+ U^{-2}(\partial U)^2g_{ij}
\end{equation}
and the Ricci scalar is 
\begin{equation}
R=-2U^{-1}\partial^2 U
\end{equation}
Substituting the Einstein tensor for the Lagrangian (\ref{masslag}) into the Einstein equations gives
\begin{eqnarray}
U^{-1}\partial^2U&=&-\mu^2(\partial f)^2\\
U^{-2}(\partial_i U)(\partial_j U)&=&
-\mu^2(\partial_i f)(\partial_j f)+GB_iB_j(1+\alpha^2 f^2)
\end{eqnarray}
These make up a complete system of equations when we include the Maxwell equation
\begin{equation}
\partial(UB)=0
\end{equation}
The second Einstein equation implies that $\partial f$, $\partial U$ and $B$ are all parallel. We therefore impose a condition $f\equiv f(u)$, $B=b(u)U^{-1}\partial U$, where
\begin{equation}
u=-\mu^{-1}\log U\label{defu}
\end{equation}
The system of equations becomes equivalent to an ordinary differential equation with independent parameter $u$,
\begin{equation}
f''+\mu\left(1+f^{\prime2}\right)f'-\alpha^2 b^2f=0\label{sos}
\end{equation}
where
\begin{equation}
b^2={1+f^{\prime2}\over 1+\alpha^2f^2}.
\end{equation}
The horizon corresponds to $u\to -\infty$ and the far region to $u\to 0$. The horizon must therefore be at a critical point of the first order system corresponding to (\ref{sos}). There is only one critical point, $(f,f')=(0,0)$, hence
\begin{equation}
(f,f')\to(0,0) \hbox{~as~} u\to-\infty.
\end{equation}
Since the critical point is a saddle, the solution is unique and exists for all values of $\mu$. Having obtained the unique solution to (\ref{sos}), we then define
\begin{equation}
V(u)=1+\mu\int_u^0b(x)e^{-\mu x}dx.\label{Vu}
\end{equation}
It is easily seen from (\ref{defu}) that
\begin{equation}
\partial_iV=V'\partial_iu=UB
\end{equation}
Hence the Maxwell equation implies $\partial^2 V=0$ and we can write 
\begin{equation}
V=1+\sum_n{GM_n\over R_n}
\end{equation}
Inverting (\ref{Vu}) gives $u(V)$.

\section{Spherically Symmetric Solutions}

In the non-extremal case we shall consider spherically symmetric metrics which can be parameterised in the form
\begin{equation}
ds^2=-{\Delta\over r^2}e^{2\delta}dt^2+{r^2\over\Delta}dr^2
+r^2d\Omega^2
\end{equation}
where $\Delta$ and $\delta$ are functions of $r$ and $t$.
After inserting the metric and the other field ansatz\"e (\ref{dirac})-(\ref{hedge}) into the Einstein field equations, one obtains 
\begin{eqnarray}
	\left(\Delta e^{\delta}f^{\prime }\right)^{\prime }
	-\frac{\lambda^{2}}{r^{2}}e^{\delta}f &=&  
	-\frac{2r^{4}}{\Delta^{3}}e^{-\delta}\dot{\Delta}\dot{f}
        -\frac{r^{4}}{\Delta}
	e^{-\delta}\dot{\delta}\dot{f}+\frac{r^{4}}{\Delta^{2}}
         e^{-\delta}\ddot{f}
	\label{sfieldeq} \\
	\delta^{\prime }  &=&  \mu ^{2}r\left(\frac{r^{4}}
	{\Delta}e^{-2\delta}\dot{f}^{2}+f^{\prime 2}\right)
	\label{dfieldeq} \\
	e^{-\delta}\left(\frac{\Delta e^{\delta}}{r}\right)^{\prime }
	 &=& 1-{\mu^2\lambda^2\over r^2}f^2-{Gp^2\over r^2}
	\label{mfieldeq} 
\end{eqnarray}
where $\mu$ and $\lambda$ are constants,
\begin{equation}
\mu^2=\pi F_\pi^2G,\qquad \lambda^2={e^4p^2\over 4\pi^3F_\pi^2}
\label{constants}
\end{equation}
The electric charge within a sphere of radius $r$ is given by equation (\ref{charge}). 

For very small $\mu$, which is the case for pions, the chiral field
has little effect on the background metric and we may take $\delta=0$ and express $ \Delta $ in terms of the mass $M$, electric charge $Q$ and magnetic charge $p$ of the black hole as
\begin{equation}
	\Delta = r^2-2GMr+G(Q^2+p^2).
\end{equation}
The skyrme field equation (\ref{sfieldeq}) on this background is therefore
\begin{equation}
	\left(\Delta f^{\prime }\right)^{\prime } - 
	{\lambda^2\over r^{2}}f=0,
	\label{simplef}  
\end{equation}
This should be solved subject to the boundary condition on 
$f_{\infty}$ which fixes the total charge,
\begin{equation}
           q_{\infty} = 
	\frac{{\rm e}^2p}{2\pi}f_{\infty} = n_B{\rm e}.
\end{equation}
The non-extremal black hole posseses two horizons at $r=r_{-}$ and 
$r=r_{+}$ $(r_{+}>r_{-})$, related to the mass and charge by
\begin{equation}
	GM=\frac{1}{2}(r_{-}+r_{+}),\qquad
        GQ^{2}=r_{-}r_{+}-Gp^2  . 
	\label{masscharge}
\end{equation}
The solution to equation (\ref{simplef}) can be obtained analytically,
\begin{equation}
f=2\pi n_B{P_q\left({r_++r_-\over r_+-r_-}
-{2r_+r_-\over  r_+-r_-}{1\over r}\right) \over
P_s\left({r_++r_-\over r_+-r_-}\right)}\label{nonex}
\end{equation}
where $P_s(z)$ is a Legendre function and
\begin{equation}
s=-{1\over 2}+\sqrt{{1\over 4}+{\lambda^2\over r_+r_-}}
\end{equation}
The black hole becomes a dyon with electric charge related to the value of $f$ at the event horizon,
\begin{equation}
Q={n_Be\over P_s\left({r_++r_-\over r_+-r_-}\right)}.
\end{equation}
This relation can be solved, together with (\ref{masscharge}),
to obtain $Q\equiv Q(M)$, showing the existence of a one parameter 
family of solutions ($n_B$ and $p$ being regarded as fixed). 
In particular, $Q\to 0$ as $M$ approaches the extremal limit $pm_{pl}$
and the meson field is expelled from the hole.

Larger values of $\mu$ may be realised for a hypothetical
model where $U$ is unrelated to pions and this is discussed below.
We will consider a static solution. We can replace $\Delta$ by a 
mass function $m(r)$, defined implicitly by the relation
\begin{equation}
\Delta=r^2-2Gmr+G(p^2+q^2)
\end{equation}
where the charge is given by equation (\ref{charge}).
The equations become
\begin{eqnarray}
m'&=&{\Delta\over 2Gr}\delta'+\mu^2\lambda^2 ff'\\
\delta'&=&\mu^2 r(f')^2\label{hatm}\\
f''&+&\left({\Delta'\over\Delta}+\delta'\right)f'
-{\lambda^2\over r^2\Delta}f=0\label{hatf}
\end{eqnarray}
Suitable boundary conditions are $f\to2\pi$ and $\delta\to 0$ as 
$r\to\infty$.

In the numerical results the fields are scaled to the horizon size,
\begin{equation}
\hat r={r\over r_+},\qquad \hat m={Gm\over r_+}
\end{equation}
The solutions are parameterised by a parameter ${\hat p}$, defined by
\begin{equation}
{\hat p}^2={Gp^2\over r_+^2}
\end{equation}
which is restricted to ${\hat p}\le 1$.

\begin{figure}
\begin{center}
\leavevmode
\epsfxsize=30pc
\epsffile{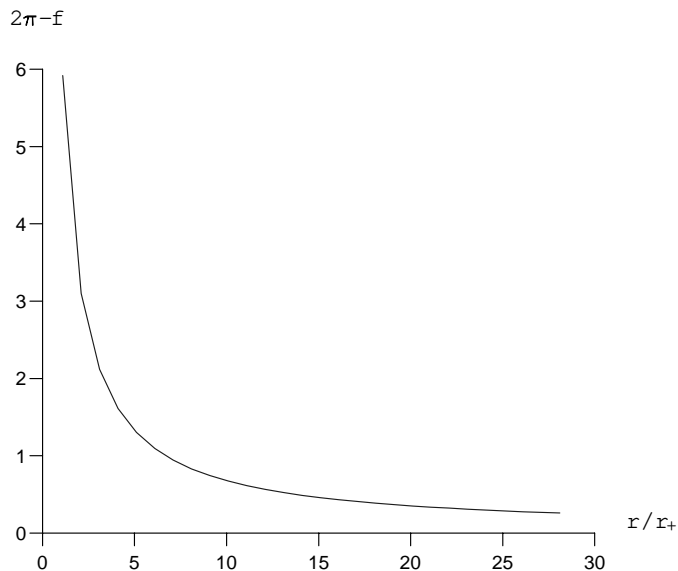}
\end{center}
\caption{Skyrme field $f$ as a function of $\hat{r}=r/r_+$ for an extremal hole and $\mu=10^{-4}$.}\protect\label{fig1}
\end{figure}

The extremal black hole solutions have $\Delta_{+}=\Delta^{\prime}_{+}=0$. The regular solution to equation (\ref{hatf}) has,
 \begin{equation}
	f_{+} = 0,\qquad Q=\frac{{\rm e}}{2\pi}f_{+}=0,\qquad{\hat p}=1.
\end{equation}    
Hence the proton lies fully outside the black hole, as we saw before.      
The numerical solution for $f$ is shown in 
fig.(\ref{fig1}). This agrees well with
the result (\ref{oneex}) of the previuos section, because the value of
$\mu$ used here is still quite small. The results are still qualitatively 
similar for chiral models with $\mu$ of order one.

For the non-extremal solution, we begin the integration of the field equations close to the horizon, with
\begin{eqnarray}
	\hat{m} &=& \hat m_0 + \hat{m}_{1}(\hat{r}-1) +
	\hat{m}_{2}(\hat{r}-1)^{2} + \cdots
	{} \\
	\delta &=& \delta_{+} + \delta_{1}(\hat{r}-1) + \cdots
	{} \\
	f &=& f_{+} + f_{1} (\hat{r}-1) + \cdots
	{}
\end{eqnarray}
where $\delta_{+}$ and $f_{+} $ are shooting parameters 
determined so as to satisfy the boundary conditions
at infinity.

\begin{figure}
\begin{center}
\leavevmode
\epsfxsize=30pc
\epsffile{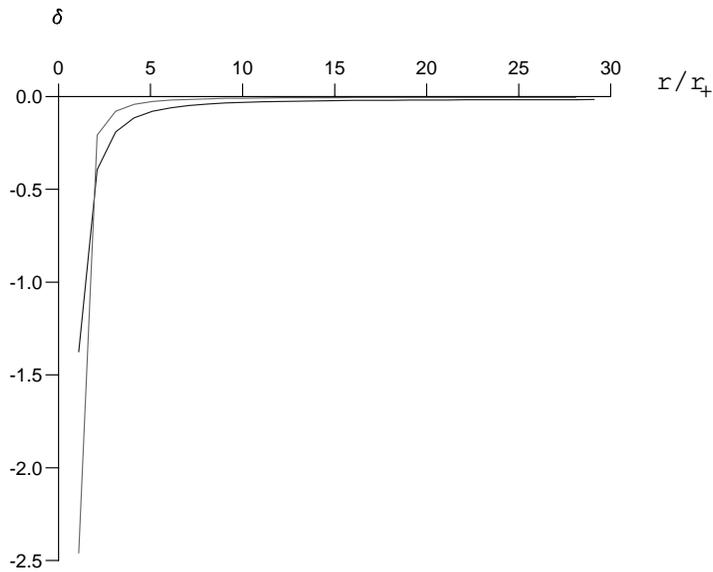}
\end{center}
\caption{Backreaction $\delta$ $(\times 10^{3})$ as a function of $\hat{r}=r/r_+$ for a non-extremal black hole, 
${\hat p} =pm_{pl}/r_+=  0.9$. Results for $\mu=10^{-3}$ and $\mu=10^{-4}$ are shown.}
	\protect\label{fig2}
\end{figure}

\begin{figure}
\begin{center}
\leavevmode
\epsfxsize=30pc
\epsffile{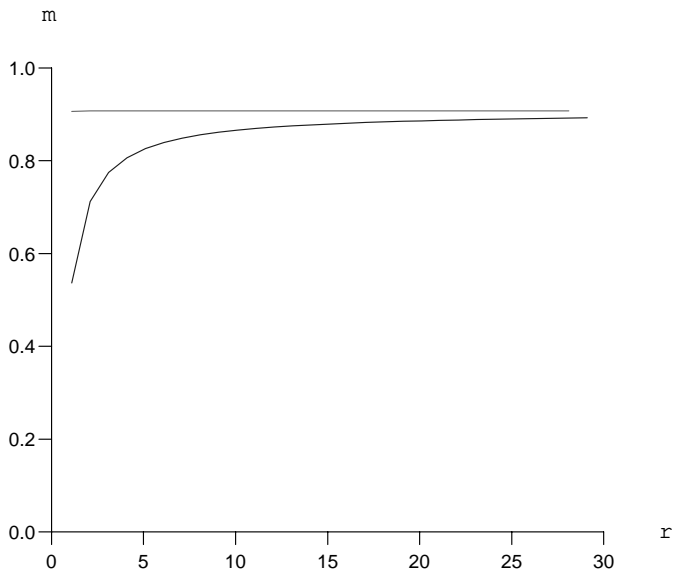}
\end{center}
\caption{Mass function $m$ as a function of $\hat{r}$ 
         for  ${\hat p} = 0.9$. Results for $\mu=10^{-3}$ and $\mu=10^{-4}$ are shown}
	\protect\label{fig3}
\end{figure}

As can be seen from the above expansion, the skyrme 
field must have a nonzero value at the horizon 
otherwise the only allowed solution is the trivial one. 
Consequently the nonextremal black hole acquires 
an electric charge 
\begin{equation}
	Q = \frac{{\rm e}}{2\pi}f_{+},
	{}
\end{equation}
and allows the skyrmion to have fractional electric charge.  
The numerical results for this solution are shown in 
fig.(\ref{fig2})-(\ref{fig3}).   
Again, these agree well with the fixed 
background for small values of $\mu$.

\begin{figure}
\begin{center}
\leavevmode
\epsfxsize=30pc
\epsffile{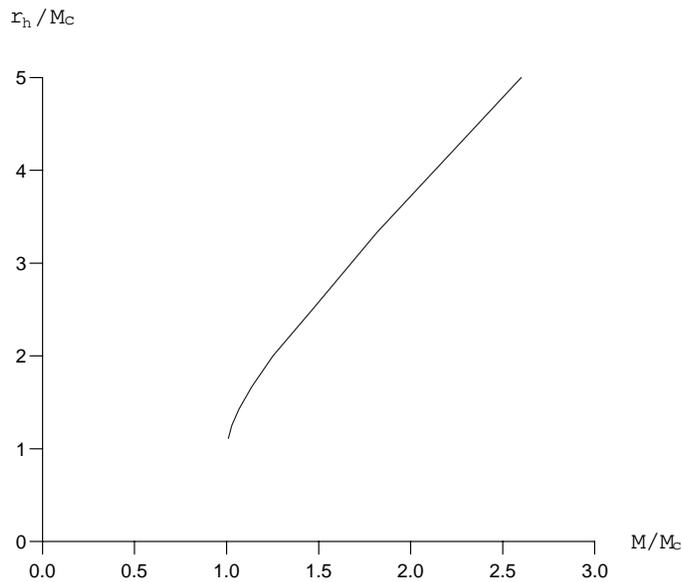}
\end{center}
\caption{Horizon radius $r_{+}/M_{c}$ as a function 
          of the black hole  mass $M/M_{c}$ for $\mu =10^{-4}$.}
	\protect\label{fig4}
\end{figure}

\begin{figure}
\begin{center}
\leavevmode
\epsfxsize=30pc
\epsffile{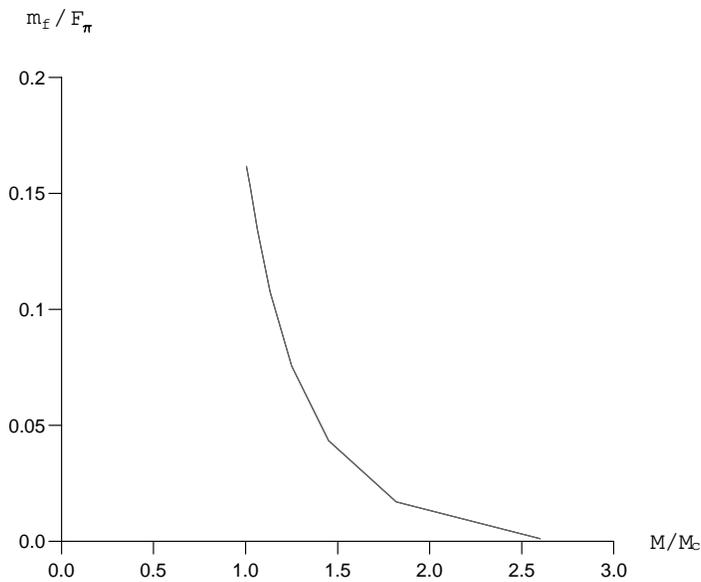}
\end{center}
\caption{Skyrmion mass $m_f$ as a function 
          of the black hole mass $M/M_{c}$ for $\mu =10^{-4}$.}
	\protect\label{fig5}
\end{figure}

We have a single one parameter family of solutions
with ${\hat p}\le 1$. In fig.(\ref{fig4})-(\ref{fig5}), we plot 
the horizon radius $r_{+}$ and 
skyrmion mass $m_{f}$ as functions of black hole mass $M$.
Figure (\ref{fig4}) is related to the entropy of the
black hole ($4\pi r_+^2$). The other figure shows how the 
proportion of the skyrmion 
which is swallowed by the black hole increases with the 
black hole mass.

\begin{figure}
\begin{center}
\leavevmode
\epsfxsize=30pc
\epsffile{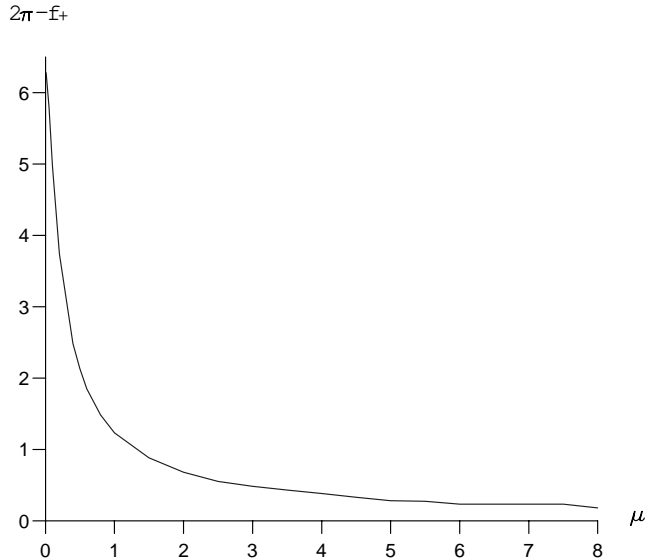}
\end{center}
\caption{The value of $f$ at the horizon for various values of $\mu$.}
	\protect\label{fig6}
\end{figure}

The last figure (\ref{fig6}) shows how the horizon value of 
$f$ changes as the coupling constant $\mu$ changes. 
Since $\mu$ characterises the mass scale of the
chiral model, this is amounts to a comparison of different
models. As can be seen from the figure, for small $\mu$ 
the electromagnetic interaction 
is dominant so that the skyrmion is absorbed 
by the black hole to a lesser extent. On the other hand 
for large $\mu$ the gravitational interaction is 
dominant so that most of the skyrmion is absorbed by 
the black hole. 

 \section{Stability Analysis}

In this section we show that the skyrmion solutions 
which we have obtained are stable under spherically
symmetric linear perturbations. We shall begin with 
the analysis of a skyrmion on the fixed background.

In the fixed background case the skyrme field is the only perturbed 
field and can be expanded about the skyrmion 
solution $f_{0}$ by writing
\begin{equation}
	f(r,t) = f_{0}(r) + e^{i\omega t}\xi(r) . 
	\label{pertf}
\end{equation}
Equation (\ref{pertf}) is inserted into equation (\ref{simplef}) 
to obtain the eigenvalue equation
\begin{equation}
	-\left(\Delta_{0}\xi^{\prime }\right)^{\prime }
         +\frac{\lambda^{2}}{r^{2}}\xi
	= \frac{r^{4}}{\Delta_{0}}\omega^{2}\xi ,
	\label{fixeigen}
\end{equation}
where the background equation has been used.

If $\omega $ is real and $\omega^{2} > 0$ the solution is stable, 
and if $\omega $ is imaginary and $\omega^{2} < 0$ 
it is unstable since the mode can grow 
or decay exponentially under the small perturbation.
To show which is the case we multiply both sides of 
equation (\ref{fixeigen}) by $\xi$ 
and integrate in $r$ from the horizon to infinity 
\begin{equation}
	\int_{r_{+}}^{\infty}\; \left[\frac{\Delta_{0}}{2} 
        \xi^{\prime 2}
	+\frac{\lambda^{2}}{r^{2}}\xi^{2}\right]\;dr = 
         \omega^{2}\int_{r_{+}}^{\infty} \; 
	\frac{r^{4}}{\Delta_{0}}\xi^{2} \; dr \; ,
	{}
\end{equation}
where integration by parts and boundary conditions have 
been used.
It can be seen that the integrands of both sides are positive definite, 
which means 
that $\omega^{2} >0$. Hence the skyrmion on the fixed background 
is linearly stable .

Next we analyse the stability of the skyrmion with backreaction.
In this case we have to expand the metric as well as the skyrme 
field around the classical solutions $f_{0}$, $ \delta_{0}$ and 
$ \Delta_{0}$
\begin{eqnarray*}
	f(r,t) & = & f_{0}(r)+f_{1}(r,t)  \\
	\delta(r,t) & = & \delta_{0}+\delta_{1}(r,t)  \\
	\Delta(r,t) & = & \Delta_{0}+\Delta_{1}(r,t)  \; .
\end{eqnarray*}
These are substituted into eq.(\ref{sfieldeq})-(\ref{mfieldeq}) 
to obtain the following 
coupled equations up to first order
\begin{eqnarray}
	\left[\left(\Delta_{0}\delta_{1}f_{0}^{\prime }
         +\Delta_{0}f_{1}^{\prime }
	+\Delta_{1}f_{0}^{\prime }\right)e^{\delta_{0}}
         \right]^{\prime }-\frac{\lambda^{2}}
         {r^{2}}\left(\delta_{1}f_{0}
	+f_{1}\right)e^{\delta_{0}}&=&
	\frac{r^{4}}{\Delta_{0}}e^{-\delta_{0}}\ddot{f}_{1}
	\label{pf} \\
	 \delta_{1}^{\prime }&=&2\mu^{2}
          rf_{0}^{\prime }f_{1}^{\prime }
	\label{del} \\
	 \left(\frac{2\mu^2\lambda^{2}}{r^{2}}f_{0}f_{1}
	 +\frac{\Delta_{0}}{r}\delta_{1}^{\prime }
           \right)e^{\delta_{0}}&=& -\left(\frac{\Delta_{1}}{r}
          e^{\delta_{0}}\right)^{\prime } .
	\label{mass}
\end{eqnarray}
Equation (\ref{mass}) can be integrated with the help 
of the static field equation,
\begin{equation}
	\Delta_{1}=-2\mu^{2}r\Delta_{0}f_{0}^{\prime }
         f_{1} \; .
	\label{intmass}
\end{equation}
Substituting equation (\ref{del}) and equation (\ref{intmass}) 
into equation (\ref{pf}) one obtains the first order equation
for $f_{1}$ 
\begin{equation}
	\left(\Delta_{0}e^{\delta_{0}}f_{1}^{\prime }
          \right)^{\prime }
	-\left[2\mu^{2}\left(r\Delta_{0}e^{\delta_{0}}
         f_{0}^{\prime 2}\right)^{\prime }
        +\frac{\lambda^{2}(M_{c})^{2}}{r^{2}}
	e^{\delta_{0}}\right]f_{1} = 
	\frac{r^{4}}{\Delta_{0}}e^{-\delta_{0}}\ddot{f}_{1} \; .
	{}
\end{equation}
Setting $f_{1}(r,t)=\xi (r)e^{i\omega t}$ one obtains an 
eigenvalue equationfor $\xi$,
\begin{equation}
	-\left(\Delta_{0}e^{\delta_{0}}
        \xi^{\prime }\right)^{\prime }
	+\left[2\mu^{2}\left(r\Delta_{0}
         e^{\delta_{0}}f_{0}^{\prime 
	2}\right)^{\prime }+\frac{\lambda^{2}
         (M_{c})^{2}}{r^{2}}
	e^{\delta_{0}}\right]\xi = \omega^{2}
	\frac{r^{4}}{\Delta_{0}}e^{-\delta_{0}}\xi \; .
	\label{backeigen}
\end{equation}
We introduce the tortoise coordinate $r^{*}$ such that
\begin{equation}
	\frac{dr^{*}}{dr} = \frac{1}{\Delta_{0}e^{\delta_{0}}}  
	{}
\end{equation}
and $r^{* }$ runs from $-\infty$ to $+\infty$ as $r$ 
runs from $r_{+}$ to $+\infty$.
Then eq.(\ref{backeigen}) is reduced to 
the Sturm-Liouville equation
\begin{equation}
	-\frac{d^{2}\xi}{dr^{* 2}} + U\xi = 
        \omega^{2}r^{4}\xi  ,
	\label{eigen}
\end{equation}
where 
\begin{equation}
	U=\left[2\mu^{2}\left(r\Delta_{0}e^{\delta_{0}}
       f_{0}^{\prime 2}\right)^{\prime }+\frac{\lambda^{2}}{r^{2}}
	e^{\delta_{0}}\right]\Delta_{0}e^{\delta_{0}}  .
	{}
\end{equation}
On the left-hand side we have 
$r^{4}$, which makes the equation different from 
the previous eigenvalue equation.
However, since $r^{4}$ remains positive through the whole space, 
the same conditions for stability as the ordinary eigenvalue 
equation can be applied.
As can be seen by examining $U$, 
$U\to 0$ as $r\to r_+$, (i.e. $r^*\to-\infty$), $U\to U_\infty>0$ as $r\to\infty$, and $U>0$ in between. 
In addition the solution does not change its shape for any 
value of the coupling constant $\mu$. 
Therefore we can safely conclude that a skyrmion 
with backreaction is also linearly stable.

 \section{Conclusion}

We are now able to describe some of the features of the
interaction between a slowly moving proton and a black hole monopole.
The free skyrmion has a magnetic moment and, if it has
the correct orientation, it will be attracted to the monopole.
When the proton approaches the black hole monopole, the fields rearrange themselves into the energetically prefered configuration of skyrmion 
hair solution described in this paper.

The skyrmion solution on the fixed black hole background 
was obtained analytically in both extremal and nonextremal cases.  
In the extremal case, the baryon number and electric 
charge are expelled by the horizon and the system 
represents protons bound to monopole black holes. We found a general
solution on the Eintein equations in the extremal case with many similarities to the BPS monopole system.

In the nonextremal case, the monopole black hole partially swallows 
the proton and transforms to a dyon black hole, 
leaving fractional electric charge and baryon number outside 
the horizon. However, Hawking radiation cannot be ignored
in this case. 

We have also obtained numerical skyrmion solutions 
in the  nonextremal case with gravitational backreaction. The effect 
of the backreaction is important only for very massive 
skyrmions i.e. very large coupling constant. The results
are qualitatively similar to the solutions without
backreaction.

Since the solutions are stable, baryon decay can only take place
when extra particle fields are included in the model. If we introduce a charged field $\phi$ on the fixed background, then following the proceedure described in reference \cite{callan}, the equations become (approximately)
\begin{eqnarray}
	\left(\Delta f^{\prime }\right)^{\prime } - 
	{\lambda^2\over r^{2}}(f-\phi)=-{r^4\over \Delta}\ddot f,\\
\left(\Delta \phi^{\prime }\right)^{\prime } +
	{\lambda^2\over r^{2}}(f-\phi)=-{r^4\over \Delta}\ddot\phi,
\end{eqnarray}
The stability arguments no longer apply. The dynamical process of a black hole swallowing a proton can be examined by solving these time-dependent 
field equations numerically.

\acknowledgments

We are grateful for discussions with K. Maeda. NS was partially supported by a British Council ORS scholarship.


\begin{references}
\bibitem{witten}Witten E, {\em Nucl.Phys.} {\bf B223} (1983) 433
\bibitem{skyrme}Skyrme T H R, {\em Proc.Roy.Soc.} {\bf A260} 
(1961) 127
\bibitem{callan}Callan C G, Jr. and Witten E, {\em Nucl.Phys.} 
{\bf B239} (1984) 161
\bibitem{moss}Luckock H and Moss I G, {\em Phys.Lett.} 
{\bf B176} (1986) 341
\bibitem{luckock}``Black Hole Skyrmions" 
Luckock H, {\em String Theory, Quantum Cosmology etc.} 
Eds. H. J. de Vega and N Sanchez (World Scientific, 1987)
\bibitem{zurich}Droz S, Heusler M and Straumann N, {\em Phys. Lett.} 
{\bf B268} (1991) 371; Bizon P and Chmaj T, {\em Phys. Lett.} {\bf B297}
(1992) 55
\bibitem{gibbons}Gibbons G W, {\em Commun. Math. Phys.} {\bf 44} (1975) 245
\bibitem{higgs}Lee K, Nair V P and Weinberg E J, 
{\em Phys. Rev. Lett.} {\bf 68} (1992) 1100; Lee K, Nair V P and Weinberg E J, {\em Phys. Rev.} {\bf D45} (1992) 2751; 
Breitenlohner P, Forgacs P and Maison D, 
{\em Nucl. Phys.} {\bf B383} (1992) 357
\bibitem{bps}Bogomol'nyi E B, {\em Sov. J. Phys.} {\bf 24} (1977) 97
\bibitem{maj}Majumbar S D {\em Phys. Rev.} {\bf 72} (1947) 390
\bibitem{pap}Papapetrou A {\em Proc. Roy. Irish. Acad.} {\bf 51} (1947) 191
\end{references}
\end{document}